\documentstyle[a4]{article}
\begin{document}
\title{Monte Carlo simulations of random copolymers at a selective interface}
\author{Gongwen Peng \thanks{Permanent address: Institute of Physics,
    Academia  Sinica, Beijing, China.}, Jens-Uwe Sommer and Alexander Blumen\\
\small
Theoretische Polymerphysik, Universit\"at Freiburg, 
Rheinstr.12,
D--79104 Freiburg, Germany}
\maketitle
\begin{abstract}
We investigate numerically using the bond--fluctuation model the
adsorption of a random AB--copolymer at the interface between two solvents.
From our results we infer several scaling relations:  
the radius of gyration  of the copolymer in the direction perpendicular to the
interface ($R_{gz}$) scales  with $\chi$, the interfacial selectivity
strength,  as
$R_{gz}=N^{\nu}f(\sqrt{N}\chi)$ where  
 $\nu$ is the usual Flory exponent and $N$ is the copolymer's length; furthermore  the monomer density at the 
interface scales as $\chi^{2\nu}$ for small $\chi$. We also determine numerically the
monomer densities in the two 
solvents and discuss their dependence on the distance from the
interface.  \\
\end{abstract}
\vspace{0cm}
\newpage

Copolymers at interfaces are very important in technical applications.
For example, 
the interface between  two immiscible
polymer melts  can be mechanically reinforced  by  
 dissolving copolymers into the system~\cite{Brown,Dai}. 
In other  applications  the surface tension between two immiscible
solvents, e.g., oil and water,  can be
reduced by adding block copolymers 
consisting of a hydrophilic and a hydrophobic part. Thus the adsorption
of copolymers at interfaces has received particular attention
[3--7]. 
For diblock copolymers,  the 
difference in the solubilities of the monomers favors the localization
of the copolymer  at the interface, with each block in
its favorable solvent. However, for $random$ copolymers frustrated
situations  may arise since the chain's connectivity forces some
monomers   to stay in 
their unfavourable solvent.
Using a Hartree-type  approach Garel $et$ $al$~\cite{Garel} have studied
the localization transition of an ideal  random chain at an interface. Yeung,
Balazs and Jasnow~\cite{Yeung} have addressed the question of
correlations in the A-- and B--distributions, a point investigated
recently through  computer
simulations by  
Balazs $et$ $al$ \cite{Balazs} in the context of  copolymer brushes.
In this paper we study numerically under excluded volume conditions  a 
single random copolymer  at an interface and   
we pay particular attention to  scaling~\cite{deGennes}.  \\

We use the bond--fluctuation model (BFM)~\cite{Kremer,Deutsch-Binder}
to perform  Monte Carlo (MC) simulations.
In the BFM the polymers obey the excluded--volume requirements, and 
the motion occurs such that the bonds do not cross each other, see 
Refs.~\cite{Kremer,Deutsch-Binder} for details. Here  we take 
 a  cubic box of size 
$L\times L\times H$, with periodic boundary conditions in the x-- and 
y--directions and  two  impenetrable 
surfaces  at $z=0$ and $z=H$. 
We study the behavior of  a single copolymer consisting of $N$
randomly chosen monomers of A-- and of B--type. We assume a symmetrical
situation: the interaction parameter of the monomers is $\chi k_BT$
when immersed in their unfavorable solvent and zero otherwise. The
solvent below the interface ($z \leq H/2$) favors A--type monomers,
and the solvent above the interface ($z \geq H/2+1$) B--type monomers.
Note that the interface is thus at $z=(H+1)/2$. 
 In
each  Monte Carlo step the chain moves by position changes of the
monomers, 
which attempt nearest neighbor steps on the underlying cubic lattice. A
move is taken into  consideration only if it satisfies the requirements of 
self--avoidance and of non--crossing of bonds. Furthermore,
energetically unfavorable moves are statistically permitted according
to the usual Boltzmann factor.  \\

We obtain results for  copolymers of lengths $N=16$,
$32$, $64$, and $128$ using systems with sizes $L=50$ and $H=100$. An
initial  configuration is generated starting with the first monomer
 near the  interface and then randomly adding the subsequent monomers
 such that self--avoidance and non--crossing of bonds are obeyed. The
 energetic aspects of the interaction with the solvents are then taken
 care of by the usual Boltzmann factor; the monomer--monomer
 interaction is only accounted for through the excluded volume
 aspects. We let the chain move  for   a long time
 according to the MC 
 prescriptions, such that the chain relaxes to equilibrium. 
 The  averaged quantities which we will show below are then obtained from
 such equilibrium configurations. We found numerically the relaxation
 time (determined using the  
autocorrelation function of the radius of gyration $R_g$ and of its
z--component $R_{gz}$ \cite{Lai}) to be around 50,000 Monte
Carlo steps (MCS), where a MCS consists of $N$ move 
attempts; we thus view the copolymer as having reached
equilibrium after 200,000 MCS. Averages are then calculated from
the configurations obtained in the subsequent  200,000 MCS. For each
copolymer length we realized 100
independent runs. \\

Figure 1 shows the normalized probability to find  monomers of A--type
and of 
B--type  at the  height $z$. In this example the copolymer length
is $N=128$ and we have chosen $\chi=3.15$. Apart from the (expected)
symmetry between the plotted curves, the figure shows that  most of
the 
monomers are located near 
the interface; the copolymer is thus adsorbed. Each of the two curves
peaks near the interface, on the favorable side. The curves decay
smoothly on the favorable side and sharply across the selective
interface. There is evidence of
frustration in the form of a secondary peak in
monomer density on the unfavorable solvent side of the interface. This
situation arises because the covalent bonding of the 
chain forces some monomers (whose neighbors are in majority of the
other type) to be immersed in the ``wrong'' solvent. This  situation
is due to an  energetic-entropic balance:  configurations with
monomers in the ``wrong'' solvent  are
energy--unfavorable but favorable in view of  entropy.   \\

Evidently, in Fig.\ 1 the asymmetry of each curve with respect to the interface  is related to the
value of $\chi$, a fact reminiscent of 
paramagnetism, where applying 
an external field increases (decreases)  the numbers of spins parallel
(antiparallel) to it. Here we determine the magnitude of the 
asymmetry across the interface by evaluating \\ 
\begin{equation}
M = \sum_{z=0}^{H} \left |{\rho_A(z)-\rho_B(z)}\right | 
  =\sum_{z=0}^{H/2} \rho_A(z) - \sum_{z=H/2+1}^{H} \rho_A(z) + 
   \sum_{z=H/2+1}^{H} \rho_B(z) - \sum_{z=0}^{H/2} \rho_B(z). 
\end{equation}
Figure 2  shows the relation between  $M$ and 
$\chi$ for different copolymer lengths $N$. Note the very small
dependence of $M$ on $N$; the curves for different $N$ almost
coincide. As in paramagnetism, 
$M$ is a linear function of $\chi$ for small $\chi$, and reaches a
constant (here unity due to  normalization) for  large $\chi$.  \\

We now turn to the question  how the copolymer behaves 
around the interface. For this we compute $R_{gz}$, the 
 z--component of the   radius of gyration of the copolymer. Here
 $R_{gz}^2=\sum_{i=1}^{N}{(z_i-\overline{z})^2}/N$, where $z_i$ is 
 the z--component of the $i$th monomer's  position  and
 $\overline{z}=\sum_{i=1}^{N}{z_i}/N$.   We are interested in the
 dependence of $R_{gz}$ on $\chi$ and hence  display in Fig.\ 3 
 $R_{gz}(\chi)/R_{gz}(0)$ as a function of $\sqrt{N}\chi$ for
different $N$. Notice that  (except for very large $\chi$) all data
collapse into a single curve. This demonstrates that $R_{gz}(\chi)$
scales with $\sqrt{N}\chi$ for small and moderately large $\chi$.
Scaling fails when $\chi$ is very large, because then the copolymers
are practically squeezed on the interface, having the A--B covalent
bonds at the interface. In the region in which the curves collapse we
observe a constant regime for very small $\chi$ followed by a
power--law decay regime for moderate $\sqrt{N}\chi$ values.  
Setting $R_{gz}(\chi)/R_{gz}(0)=f(\sqrt{N}\chi)$ with 
$f(y)=1$  for  $y < y_c$ 
and 
$f(y)=y^{-\alpha}$  for $y > y_c$, we find from Fig.\ 3 that  
 the crossover value $y_c$ is roughly $y_c \approx 10$. Furthermore,
 from a best fit to 
the data in the power--law  regime we obtain numerically for the  
exponent $\alpha$ that  $\alpha=1.112 
\pm 0.10$ i.e., $\alpha/2=0.56 \pm 0.05$.  
This result is in agreement with   the scaling arguments of
Refs.~\cite{Sommer-Daoud,Jens}, which predict that  $\alpha$ should
equal $2\nu$,  where $\nu$ is the usual Flory exponent, $\nu \approx
0.588$ for excluded--volume  chains. 
 Scaling with $\sqrt{N}\chi$ is consistent
with the ``blob''--picture of a random copolymer at an interface
\cite{Sommer-Daoud,Jens}. A blob is a chain segment containing $g$ ($g
\gg 1$) monomers, and thus has roughly $g^{1/2}$ monomers of one type in
excess; the total number of blobs is $N/g$ and  their number at the
interface  depends on
$\sqrt{N}\chi$~\cite{Jens}.  \\

Another example of scaling is provided by the monomer  density exactly at the
interface $z=(H+1)/2$. In our (discrete) lattice model we thus compute 
\begin{equation}
\rho_s
=\frac{1}{2}[\rho_A(\frac{H}{2})
+\rho_B(\frac{H}{2}+1)]~~.
\end{equation}
In Fig.\ 4 we display in a log--log plot $\rho_s$ as a function of
$\chi$ for different polymer lengths $N$. Again the data
for different $N$ coincide, possibly with the exception of the very
small $\chi$  regime; here, however, $\rho_s$ is small and thus the
relative error is rather large.  For very 
large $\chi$, $\rho_s$ gets to be  independent of $\chi$,  as the
chains are then  
 squeezed on the surface.  In the
moderate $\chi$ regime $\rho_s$ scales with $\chi$ as a power--law
$\rho_s \sim \chi^{\beta}$,  
which is illustrated by the linearity of the data in the log--log plot
of Fig.\ 4. From a best fit to the data in this  regime we
obtain numerically that the  exponent $\beta$ is $1.14 \pm 0.06$. 
This result  can be compared to the expression $\beta=2\nu$ in  the
``blob'' picture  
\cite{Sommer-Daoud,Jens}, i.e., taking the accepted $\nu\approx 0.588$  for
excluded volume chains to $\beta \approx 1.176$. The derivation of
$\beta=2\nu$ 
according to  
Refs.~\cite{Sommer-Daoud} and \cite{Jens} starts from  the scaling of
physical variables with  $N^{1/2}\chi$. Hence  the
total number of monomers at the interface obeys 
$N_s=N^{1-\nu}f(N^{1/2}\chi)$, where $f(0)=1$. Noting that for large
$\chi$ $N_s$ is proportional to 
 $N$ requires that $f(y) \sim y^m$, with $1-\nu+m/2 =
1$. Hence $m = 2\nu$ and  $N_s \sim N\chi^{2\nu}$. Due to
normalization one finds $\rho_s = N_s/N \sim 
\chi^{2\nu}$.  
  \\

We conclude by showing that even the total monomer density 
scales with $\chi^{2\nu}$. In Fig.\  5(a) we plot
$(\rho_A(z)+\rho_B(z))/\rho_0$ as a function of  $(z-z_0) \chi^{2\nu}$
with $z_0=(H+1)/2$ and 
$\rho_0=\frac{1}{2}[\rho_A(\frac{H}{2})+\rho_B(\frac{H}{2})+
\rho_A(\frac{H}{2}+1)
+\rho_B(\frac{H}{2}+1)])$. For a series of $N$ and $\chi$ values we
find that all data collapse to a single 
curve  when $\chi$ lies in the power--law regime of Fig.\ 3. This
implies the scaling relation 
\begin{equation}
(\rho_A(z)+\rho_B(z)) =
\rho_0 g((z-z_0) \chi^{2\nu})~~.
\end{equation}
Eq.\ (3) does not hold when $\chi$ lies outside the power--law regime. 
Fig.\ 5(b) displays the data of Fig.\ 5(a) in a semi--logarithmic
plot. The appearance of two almost straight  lines rules out a
Gaussian behavior and 
suggests to approximate the wings of $g(y)$ by an 
exponential  form. We find for the wings $g(y) \sim \exp{(-\gamma|y|)}$,
with $\gamma = 0.10 \pm 0.02$.
\\

In conclusion, we have studied numerically the adsorption
of random copolymers at a selective interface. The main results are as
follows. For small and for moderately large $\chi$ the 
magnitude $M$ of  the asymmetry, as defined in Eq.\ (1), grows
linearly with $\chi$ and is independent of $N$. Furthermore, $R_{gz}$, 
the radius of 
gyration in the direction perpendicular to the interface scales  as
$R_{gz}=N^{\nu}f(\sqrt{N}\chi)$.  
The monomer density $\rho_s$  at the
interface, Eq.\ (3), scales  as $\chi^{2\nu}$; moreover, 
 for $\chi$ in the power--law regime the total monomer density obeys 
$\rho_A(z)+\rho_B(z) =
\rho_0 g(|z-z_0| \chi^{2\nu})$  where
 $g(y)$ is close to being exponential. \\

This work was supported by the Deutsche Forschungsgemeinschaft (SFB
60),  by the Fonds der Chemischen Industrie and by PROCOPE,
administrated by the DAAD. GP thanks the Alexander
von Humboldt 
Foundation for a fellowship. \\

\newpage
\noindent
{\bf Figure Captions}\\
Figure 1: The probability (normalized density) $\rho$ to find a
monomer of A--type (diamonds) or B--type (crosses) at height $z$. Here the
copolymer length is $N=128$ and 
$\chi=3.15$. \\

Figure 2: The dependence of the asymmetry parameter $M$, Eq.(1) on
$\chi$  for different polymer lengths. \\

Figure 3: $R_{gz}(\chi)/R_{gz}(0)$ as a function of the  scaling variable
  $\sqrt{N}\chi$ for different polymer lengths. \\

Figure 4: The density $\rho_s$ exactly at the interface  versus the
interfacial selectivity strength $\chi$ for different polymer lengths.
\\

Figure 5(a): The total monomer density plotted as a function of
$(z-z_0)\chi^{2\nu}$, see text for details. The parameter values are 
$a$: $N=32$, $\chi=3.15$; $b$: $N=64$, $\chi=2.20$; $c$: $N=64$,
$\chi=3.15$; $d$: $N=128$, $\chi=1.55$; $e$: $N=128$, $\chi=2.30$;
$f$: $N=128$, $\chi=3.15$. 

Figure 5(b): Same as (a) but plotted semi--logarithmically. \\

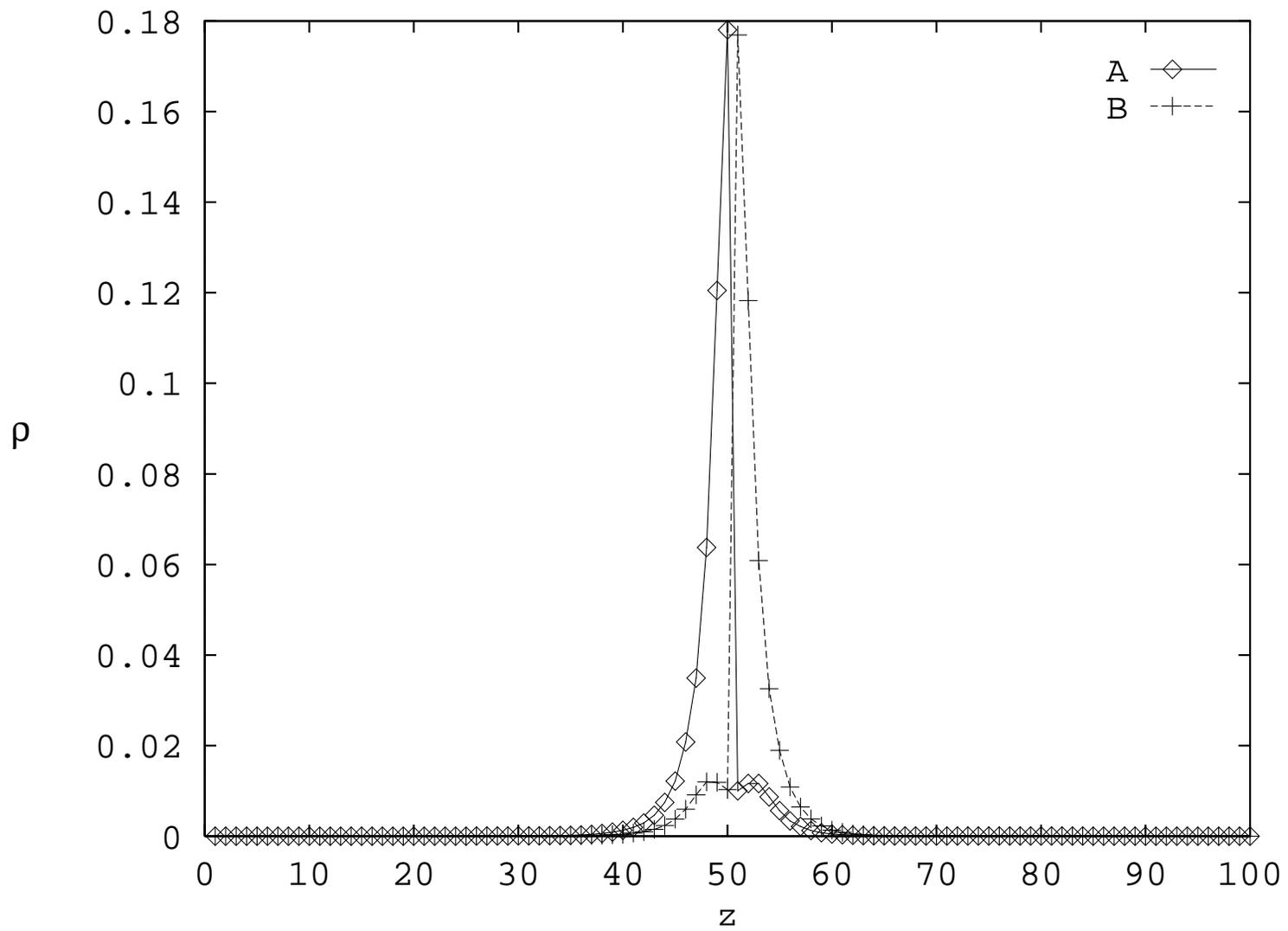
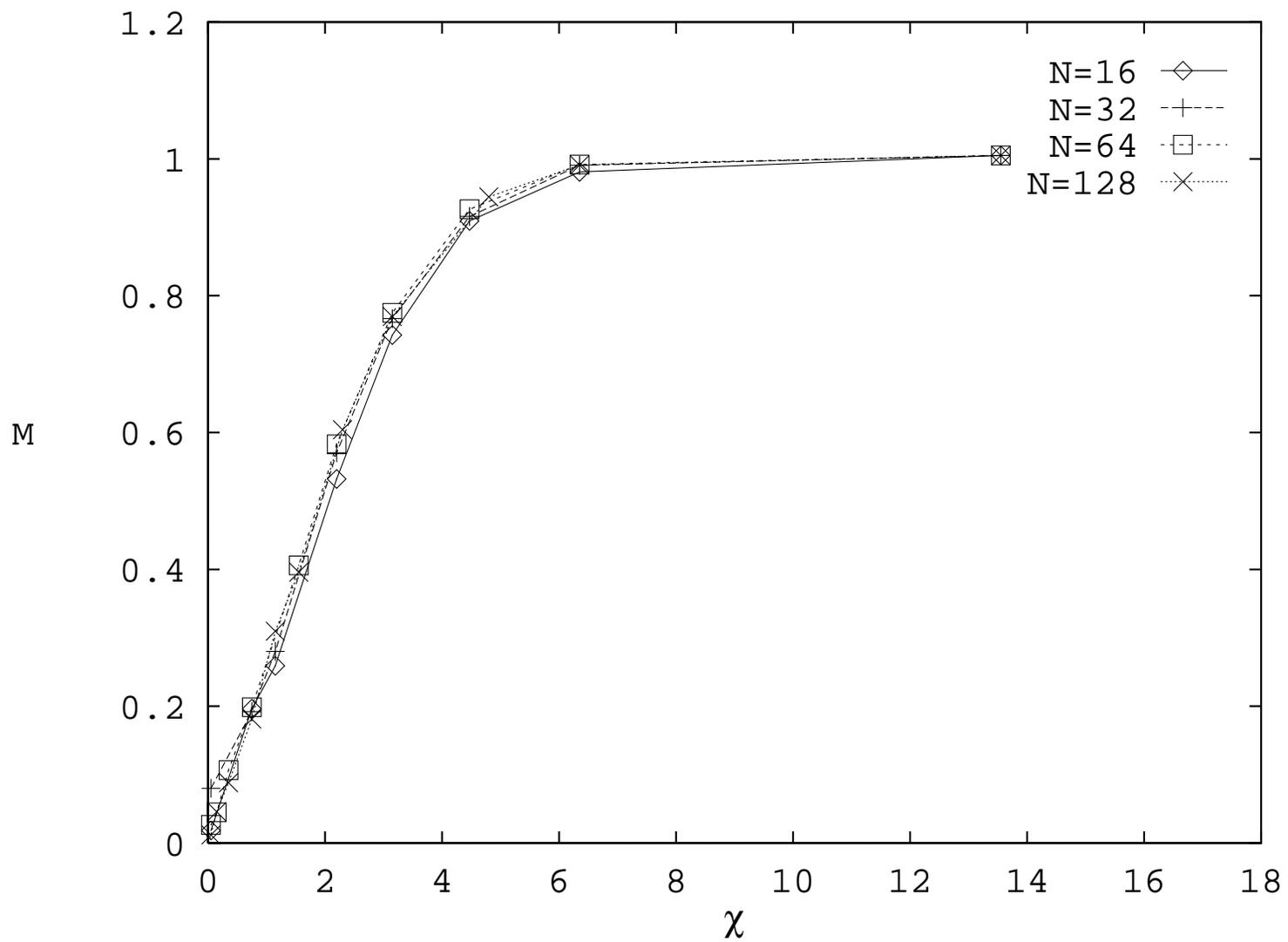
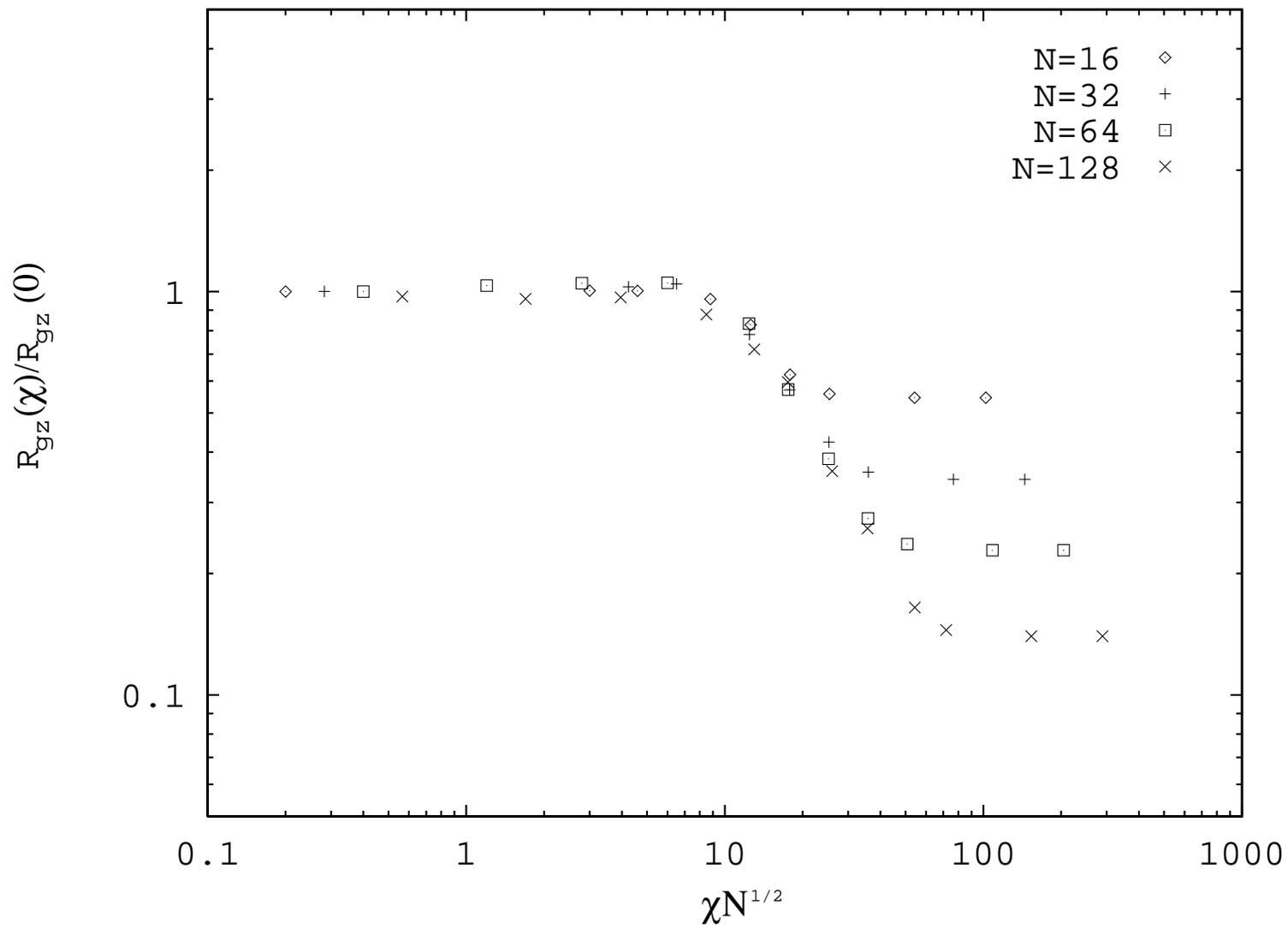
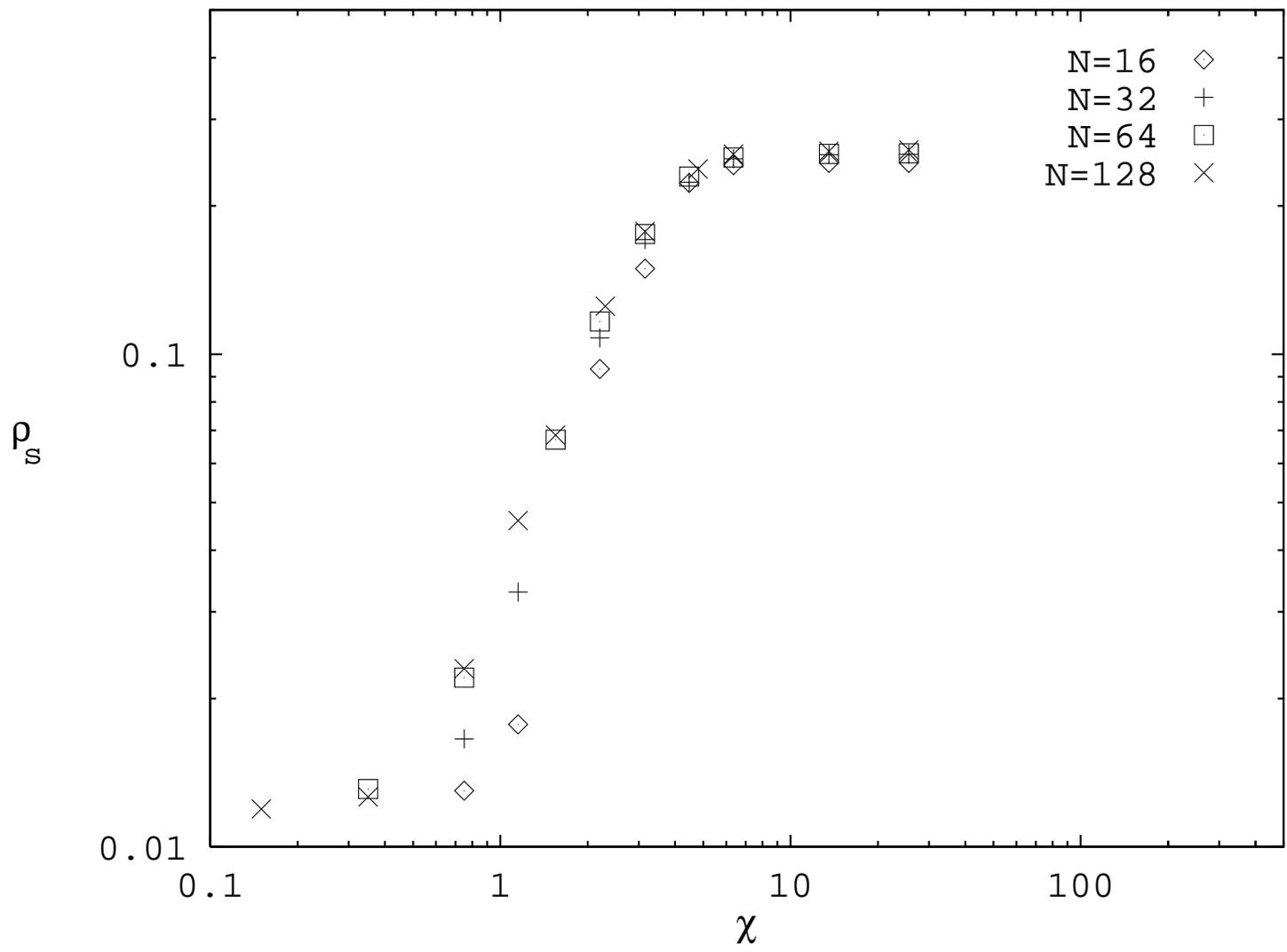
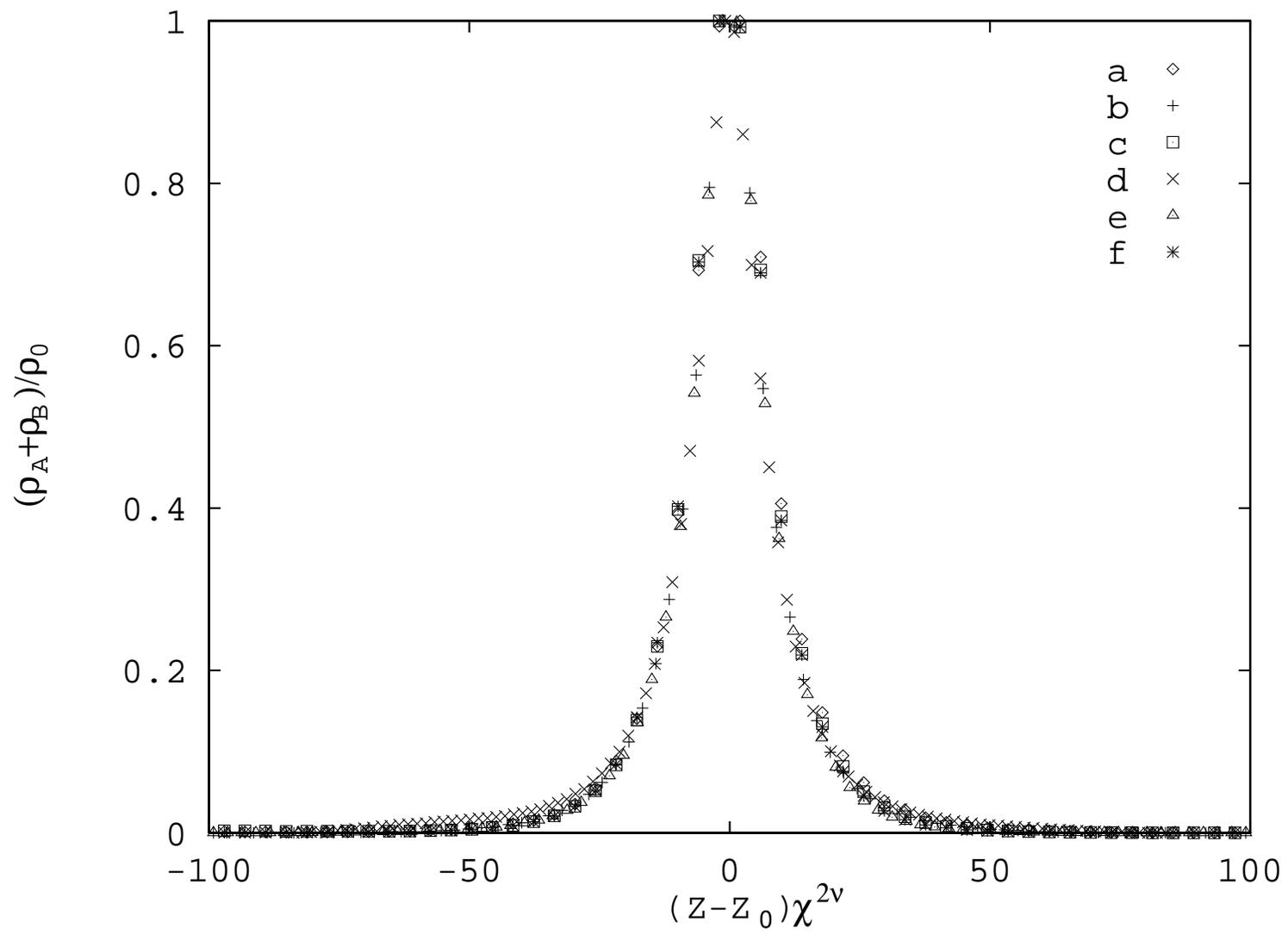
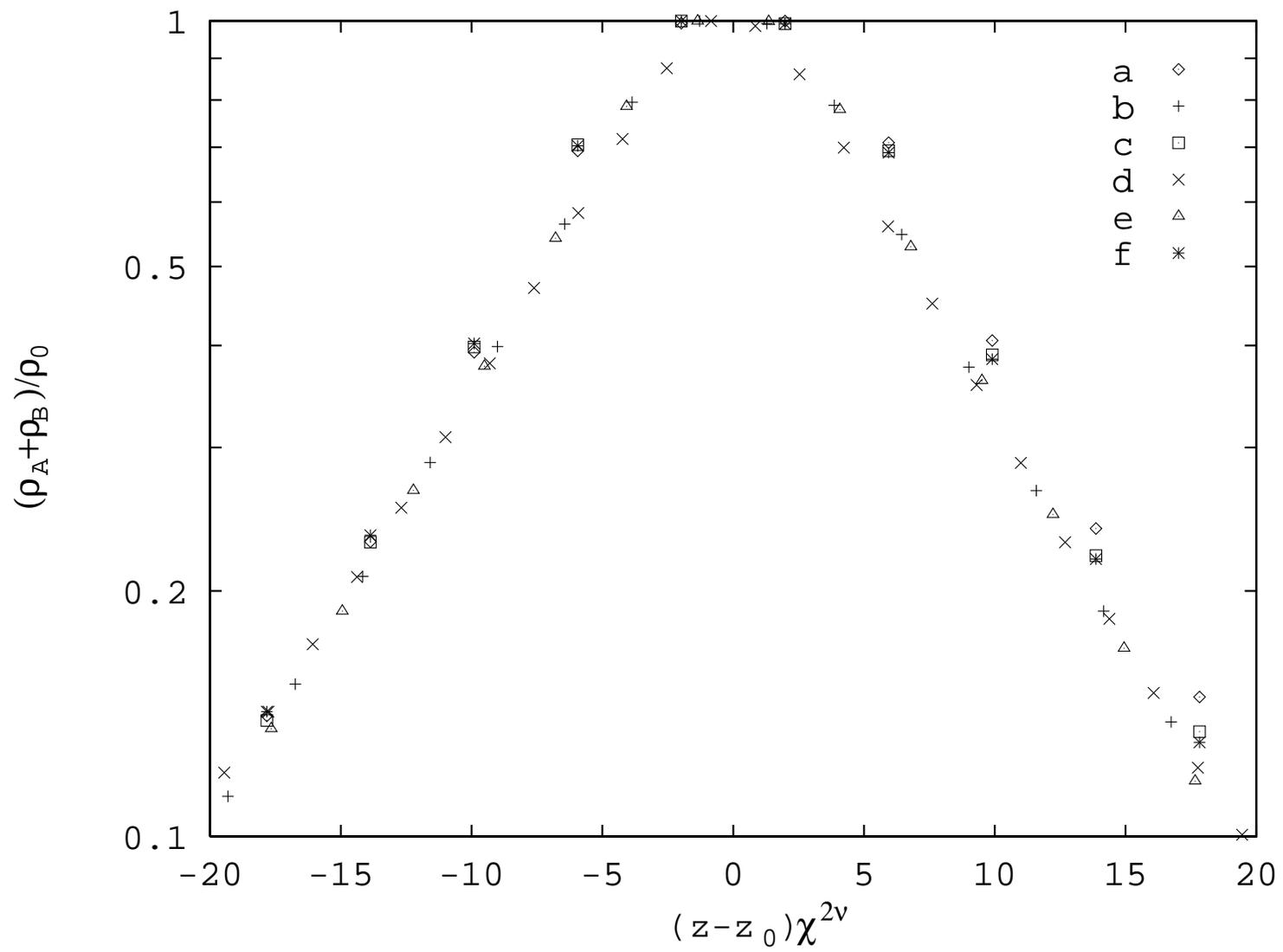
\end{document}